\documentstyle[prd,aps]{revtex}
\begin{document}
\draft
\preprint{gr-qc/0009077}
\title{The Bardeen Model as a Nonlinear Magnetic Monopole}
\author{Eloy Ay\'on--Beato and Alberto Garc\'{\i}a}
\address{Departamento~de~F\'{\i}sica,~Centro~de~Investigaci%
\'on~y~de~Estudios~Avanzados~del~IPN\\
Apdo. Postal 14-740, 07000 M\'exico DF, MEXICO}
\maketitle

\begin{abstract}
The Bardeen model---the first regular black hole {\em model\/} in General
Relativity---is reinterpreted as the gravitational field of a nonlinear
magnetic monopole, i.e., as a magnetic solution to Einstein equations
coupled to a nonlinear electrodynamics.
\end{abstract}


The study on the global regularity of black hole solutions is quite
important in order to understand the final state of gravitational collapse
of initially regular configurations. The Penrose (weak) cosmic censorship
conjecture claims that the singularities predicted by the celebrated
``singularity theorems'' \cite{H-E,Senov98} must be dressed by event
horizons (cf. \cite{Wald97,JMagli99} for recent reviews). The first black
hole solutions known in General Relativity were all in accordance with such
point of view. However, the conjecture does not forbid the existence of
regular (singularity--free) black holes. In fact, as it was pointed by
Hawking and Ellis \cite{H-E}, and more recently by Borde \cite
{Borde94,Borde97}, the model proposed by Bardeen \cite{Bardeen68} at the
early stage of investigations on singularities, explicitly shows the
impossibility of proving more general singularity theorems, if one avoids
the assumption of either the strong energy condition or the existence of
global hyperbolicity. The Bardeen model \cite{Bardeen68} is a regular black
hole space--time satisfying the weak energy condition. This model has been
recently revived by Borde \cite{Borde94,Borde97}, who also clarified the
avoidance of singularities in this space--time, and in similar examples, as
due to a topology change in the evolution of the space--like slices from
open to closed \cite{Borde97}.

Other regular black hole models were proposed later \cite
{Borde94,BarrFrolov96,MMPSenovilla96,CaboAyon97}. None of these ``Bardeen
black holes,'' as they have been called by Borde \cite{Borde97}, is an exact
solution to Einstein equations, thus, there is no known physical sources
associated with any of them. Regular black hole solutions to Einstein
equations with physically reasonable sources has been reported only recently 
\cite{AyonGarcia98,AyonGarcia99a,AyonGarcia99b,Magli97}. The objective of
this contribution is to provide the pioneering Bardeen model \cite{Bardeen68}
with a physical interpretation. The Bardeen model is described by the metric 
\begin{equation}
\text{\boldmath$g$}=-\left( 1-\frac{2mr^2}{(r^2+g^2)^{3/2}}\right) \text{%
\boldmath$dt$}^2+\left( 1-\frac{2mr^2}{(r^2+g^2)^{3/2}}\right) ^{-1}\text{%
\boldmath$dr$}^2+r^2\text{\boldmath$d\Omega $}^2.  \label{eq:Bardeen}
\end{equation}
It can be noted that this metric asymptotically behaves as $%
-g_{tt}=1-2m/r+3mg^2/r^3+O(1/r^5)$; from the $1/r$ term it follows that the
parameter $m$ is associated with the mass of the configuration
(incidentally, this fact can also be verified from the explicit evaluation
of the ADM mass definition), but, the following term goes as $1/r^3$,
therefore this does not allow one to associate the parameter $g$ with some
kind of ``Coulomb'' charge as, for instance, in the Reissner--Nordstr\"om
solution. This fact causes that up--to--date there is no known physical
interpretation for the regularizing parameter $g$. The main objective of
this work is to show that $g$ is the monopole charge of a self--gravitating
magnetic field described by a nonlinear electrodynamics.

The Bardeen model (\ref{eq:Bardeen}) describes a regular space--time; this
can be realized from the analytical expressions of its curvature invariants 
\begin{equation}
R=\frac{6mg^2\left( 4\,g^2-r^2\right) }{\left( r^2+g^2\right) ^{7/2}},
\label{eq:R}
\end{equation}
\begin{equation}
R_{\mu \nu }R^{\mu \nu }=\frac{18m^2g^4\left(
8\,g^4-4\,g^2r^2+13\,r^4\right) }{\left( r^2+g^2\right) ^7},
\label{eq:Ricci}
\end{equation}
\begin{equation}
R_{\mu \nu \alpha \beta }R^{\mu \nu \alpha \beta }=\frac{12m^2\left(
8\,g^8-4\,g^6r^2+47\,g^4r^4-12\,g^2r^6+4\,r^8\right) }{\left( r^2+g^2\right)
^7},  \label{eq:Riemann}
\end{equation}
which are all regular everywhere. For certain range of the parameter $g$ the
Bardeen metric describes also a black hole. Making the substitutions $%
y=r/|g| $, $s=|g|/2m$ we rewrite $g_{tt}$ as 
\begin{equation}
-g_{tt}=A(x,s)\equiv 1-\frac 1s\frac{x^2}{(1+x^2)^{3/2}}.  \label{eq:A}
\end{equation}
As it can be noted from the derivative of the last function, it has a single
minimum at $x_{{\rm {m}}}=\sqrt{2}$, independently of the nonvanishing value
of $s$. The equation $A(x_{{\rm {m}}},s)=0$ is solved by the single root $s_{%
{\rm {c}}}=2/\sqrt{27}$. At $x_{{\rm {m}}}$, for $s<s_{{\rm {c}}}$ the
quoted minimum of $A(x,s)$ is negative, for $s=s_{{\rm {c}}}$ the minimum
vanishes and for $s>s_{{\rm {c}}}$ the minimum is positive. From the
regularity of the curvature invariants (\ref{eq:R}--\ref{eq:Riemann}), it
follows that for $s{\leq }s_{{\rm {c}}}$ the singularities appearing in (\ref
{eq:Bardeen}), due to the vanishing of $A$, are only
coordinate--singularities describing the existence of event horizons.
Consequently, we are in the presence of black hole space--times for $g^2\leq
4s_{{\rm {c}}}^2m^2=(16/27)m^2$. For the strict inequality $g^2<(16/27)m^2$,
there are inner and event horizons for the Killing field $\text{\boldmath$k$}%
=\text{\boldmath$\partial /\partial {t}$}$, defined by $k_\mu k^\mu
=g_{tt}=0 $. For the equality $g^2=(16/27)m^2$, the horizons shrink into a
single one, where also $\nabla _\nu (k_\mu k^\mu )=0$, {\em i.e.\/}, this
case corresponds to an extreme black hole as in the Reissner--Nordstr\"om
solution. The extension of the Bardeen metric beyond the horizons, up to $%
r=0 $, becomes apparent by passing to the standard advanced and retarded
Eddington--Finkelstein coordinates, in terms of which the metric is
well--behaved everywhere, even in the extreme case. The maximal extension of
the Bardeen metric can be achieved by following the main lines used for the
Reissner--Nordstr\"om solution, taking care of course, of the more involved
integration in the present case of the tortoise coordinate $r^{*}\equiv \int
A^{-1}dr$. The global structure of the Bardeen space--time is similar to the
structure of the Reissner--Nordstr\"om black hole, as it has been pointed
out by previous authors \cite{H-E,Borde97}, except that the usual
singularity of the Reissner--Nordstr\"om solution, at $r=0$, is smoothed out
and now it simply corresponds to the origin of the spherical coordinates.
This fact implies that the topology of the slices change from $S^2\times 
{\rm {I\!R}}$ outside the black hole to $S^3$ inside of it, as it has been
pointed by Borde \cite{Borde97}.

In what follows we will show that it is possible to provide a physical
interpretation to the quoted parameter $g$, and to think of the Bardeen
model as a regular black hole solution of General Relativity simply by
coupling to the Einstein equations an appropriate nonlinear electrodynamics.
In this context $g$ is the monopole charge of a magnetic field ruled by the
quoted nonlinear electrodynamics. The dynamics of the proposed theory is
governed by the action 
\begin{equation}
{\cal S}=\int dv\,\left( \frac 1{16\pi }R-\frac 1{4\pi }{\cal L}(F)\right) ,
\label{eq:action}
\end{equation}
where $R$ is scalar curvature, and ${\cal L}$ is a function of $F\equiv 
\frac 14F_{\mu \nu }F^{\mu \nu }$, where $F_{\mu \nu }=2\nabla _{[\mu
}A_{\nu ]}$ is the electromagnetic strength. We would like to recall that
there are more general Lagrangians depending also on the second invariant, $%
F_{\mu \nu }^{*}F^{\mu \nu }$, but for the objective of this work it is
enough to consider only an action as the one given in (\ref{eq:action}). The
Einstein--nonlinear--electrodynamics field equations resulting from action (%
\ref{eq:action}) are 
\begin{equation}
G_\mu ^{~\nu }=2({\cal L}_FF_{\mu \lambda }F^{\nu \lambda }-\delta _\mu
^{~\nu }{\cal L}),  \label{eq:Ein}
\end{equation}
\begin{equation}
\nabla _\mu \left( {\cal L}_FF^{\alpha \mu }\right) =0.  \label{eq:Max}
\end{equation}
where, ${\cal L}_F\equiv \partial {\cal L}/\partial F$. The particular
non--linear electrodynamics source used to derive the Bardeen black hole is
determined by the following function ${\cal L}$: 
\begin{equation}
{\cal L}(F)=\frac 3{2sg^2}\left( \frac{\sqrt{2\,g^2F}}{1+\sqrt{2\,g^2F}}%
\right) ^{5/2},  \label{eq:L}
\end{equation}
where $s$ stands for $s\equiv |g|/2m$; $g$ and $m$ are free parameters which
we anticipate to be associated with magnetic charge and mass respectively.
To obtain a solution compatible with (\ref{eq:Bardeen}), we consider a
static and spherically symmetric configuration 
\begin{equation}
\text{\boldmath$g$}=-\left( 1-\frac{2M(r)}r\right) \text{\boldmath$dt$}%
^2+\left( 1-\frac{2M(r)}r\right) ^{-1}\text{\boldmath$dr$}^2+r^2\text{%
\boldmath$d\Omega $}^2,  \label{eq:spher}
\end{equation}
and assume the following magnetic ansatz for the Maxwell field 
\begin{equation}
F_{\mu \nu }=2\delta _{[\mu }^\theta \delta _{\nu ]}^\varphi B(r,\theta ).
\label{eq:stat}
\end{equation}
With these choices, equations (\ref{eq:Max}) are easily integrated, 
\begin{equation}
F_{\mu \nu }=2\delta _{[\mu }^\theta \delta _{\nu ]}^\varphi f(r)\sin
(\theta ).  \label{eq:dielec}
\end{equation}
Using now that 
\begin{equation}
0=\text{\boldmath$dF$}=f^{\prime }(r)\sin (\theta )\text{\boldmath$dr$}%
\wedge \text{\boldmath$d\theta $}\wedge \text{\boldmath$d\varphi $},
\label{eq:dF}
\end{equation}
we conclude that $f(r)=\text{const.}=g$, where the integration constant has
been chosen as $g$. As it was anticipated above, $g$ is the magnetic
monopole charge of the configuration: 
\begin{equation}
\frac 1{4\pi }\int_{S^\infty }\text{\boldmath$F$}=\frac g{4\pi }\int_0^\pi
\int_0^{2\pi }\sin (\theta )d\theta d\text{$\varphi =g$},  \label{eq:monop}
\end{equation}
where $S^\infty $ is a sphere at infinity.

The $_t^{~t}$ component of Einstein equations (\ref{eq:Ein}) yields 
\begin{equation}
M^{\prime }(r)=r^2{\cal L}(F).  \label{eq:tt}
\end{equation}
Substituting ${\cal L}$ from (\ref{eq:L}) with $F=g^2/2r^4$, and using that $%
m=\lim_{r\rightarrow \infty }M(r)$, one can write the integral of (\ref
{eq:tt}) as 
\begin{equation}
M(r)=m-3mg^2\int_r^\infty dy\frac{y^2}{\left( y^2+g^2\right) ^{5/2}}.
\label{eq:int}
\end{equation}
The last integral can be easily accomplished using hyperbolic functions,
yielding 
\begin{equation}
M(r)=\frac{mr^3}{\left( r^2+g^2\right) ^{3/2}}.  \label{eq:Q}
\end{equation}
Substituting $M(r)$ into (\ref{eq:spher}) one finally obtains the Bardeen
metric (\ref{eq:Bardeen}).

It can be noted that the proposed source (\ref{eq:L}) satisfies the weak
energy condition. Let {\boldmath$X$} be a time--like field, without loss of
generality it can be chosen normal ($X_\mu X^\mu =-1$). Using the right hand
side of (\ref{eq:Ein}), one can write the local energy density along {%
\boldmath$X$} as 
\begin{equation}
4\pi T_{\mu \nu }X^\mu X^\nu ={\cal L}+E_\lambda E^\lambda {\cal L}_F,
\label{eq:wec}
\end{equation}
where $E_\lambda \equiv F_{\lambda \mu }X^\mu $ is by definition orthogonal
to {\boldmath$X$}, so it is an space--like vector ($E_\lambda E^\lambda >0$%
). It follows from (\ref{eq:wec}) that if ${\cal L}\ge 0$ and ${\cal L}_F\ge
0$ the local energy density along any time--like field is non--negative
everywhere, which is the requirement of the weak energy condition. For the
proposed non--linear electrodynamics the non--negativeness of these
quantities follows from the definition (\ref{eq:L}), hence, the matter
proposed by us as source of the Bardeen model satisfies the weak energy
condition.

Summarizing, for the Bardeen metric (\ref{eq:Bardeen}) we found a field
source related to a nonlinear electrodynamics given by the Lagrangian (\ref
{eq:L}). The solution to the Einstein equations coupled with the
energy--momentum tensor associated to the magnetic strength (\ref{eq:dielec}%
) corresponds to a self--gravitating magnetic monopole charge $g$ (\ref
{eq:monop}). It is well known the Bardeen model, now climbed to the status
of an exact solution of Einstein equations, fulfills the weak energy
condition and is regular everywhere, although the invariant of the
associated electromagnetic field exhibits the usual singular behavior $%
F=g^2/2r^4$ of magnetic monopoles. Finally, we would like to point out that
the nonlinear electrodynamics used is stronger than the Maxwell one, in the
weak field limit, as it can be seen from the expansion of the Lagrangian (%
\ref{eq:L}); ${\cal L}(F\ll 1)\sim FF^{1/4}$, while for Maxwell
electrodynamics ${\cal L}_{\text{M}}(F)=F$.

\acknowledgments
This work was partially supported by the CONACyT Grant 32138E and the
Sistema Nacional de Investigadores (SNI). EAB also thanks all the
encouragement and guide provided by his recently late father: Erasmo Ay\'on
Alayo.

\end{document}